\documentstyle[preprint,eqsecnum,aps]{revtex}
\tighten
\begin{document}
\draft

\title{Semiclassical models for uniform-density
Cosmic Strings and Relativistic Stars}

\author{Manuela CAMPANELLI\thanks{Electronic Address:
Manuela.Campanelli@uni-konstanz.de}}
\address{Institut f\"ur Theoretische Physik der Universit\"at Bern\\
Sidlerstra\ss e 5, CH-3012 Bern, Switzerland\\
and\\
Fakult\"at f\"ur Physik der Universit\"at Konstanz\\
Postfach 5560 M 674, D-78434 Konstanz, Germany}
\author{Carlos O. LOUSTO\thanks{Electronic Address:
lousto@mail.physics.utah.edu}}
\address{Department of Physics, University of Utah\\
201 JBF, Salt Lake City, UT 84112, USA}
\date{\today}
\maketitle

\begin{abstract}

In this paper we show how quantum corrections, although
perturbatively small, may play an important role in the
analysis of the existence of some classical models.
This, in fact, appears to be the case of static,
uniform--density models of the interior metric of
cosmic strings and neutron stars.
We consider the fourth order semiclassical equations
and first look for perturbative solutions in the coupling
constants $\alpha$ and $\beta$
of the quadratic curvature terms in the effective
gravitational Lagrangian. We find that there is not a
consistent solution; neither for strings nor for spherical
stars. We then look for non--perturbative solutions and
find an explicit approximate metric for the case of straight
cosmic strings.
We finally analyse the contribution of the non--local
terms to the renormalized energy--momentum tensor
and the possibility of this terms to allow for a
perturbative solution. We explicitly build up a particular
renormalized energy--momentum tensor to fulfill that end.
These state--dependent corrections are found by simple
considerations of symmetry, conservation law and trace
anomaly, and are chosen to compensate for the local terms.
However, they are not only ad hoc, but have to depend on
$\alpha$ and $\beta$, what is not expected to first
perturbative order. We then conclude that non--perturbative
solutions are valuable for describing certain physical
situations.

\end{abstract}
\pacs{04.50.+h,11.27.+d,97.60.Jd}

\section{Introduction}

It is well--known that semiclassical quantization of matter
fields in a curved background reveals the necessity
to include higher order terms in the gravitational
Lagrangian in order to absorb the divergent behavior
coming from the one--loop expansion of the matter field
(see Ref.\ \cite{BD82} for a review). These terms
can be expressed as quadratic combinations of the curvature
tensor and of its derivatives. The resulting theory has field
equations including fourth order
derivatives of the metric.

The loop expansion to quantum gravity can be
formally interpreted as an expansion in powers of $l^2_{pl}$
(i. e. in powers of $\hbar$ in $G=c=1$ units).
One formally expands both a quantum metric operator 
$\hat{g}_{\mu\nu}=\eta_{\mu\nu}+\hat{\Psi}_{\mu\nu}$
and a quantum matter field operator $\hat{\Phi}_{\mu\nu}$
(for simplicity a scalar field) about any fixed vacuum
classical solution $g^{(0)}_{\mu\nu}$.
However, if we consider the full series in this expansion we
discover that not only
does not converge but is non--renormalizable.
In the lowest (one--loop) order, the matter divergences
can be absorbed in the constants of the theory.
However, this finiteness cannot be attained at this 
one--loop order for gravity coupled to matter fields.
This raises the well--known graviton loop problem since,
there is no obvious mathematical nor physical
justification for dropping the `one--loop' of the
graviton, because in this scheme the one--loop quantum
effects of the metric
are just as important as those of the matter field.

On the other hand, the alternative $1/N$ approach to
semiclassical gravity was investigated in Ref.\ \cite{T77}.
In this approach one considers the Einstein--Hilbert
gravitational action plus matter\ (\ref{I.1}), where
in the matter Lagrangian ${\cal L}_m^N$ one
assumes the presence of $N$ identical and free
(i. e. non-self interacting) conformally invariant
scalar fields, all of which are in the same quantum state.
Then, instead of expanding the theory in powers of
$\hbar$, which results in the ordinary loop expansion,
one expands in powers of $1/N$.
In this expansion each field is now coupled to gravity with coupling
proportional to $1/N$ and all matter contributions (classical and
one--loop) are included already at the leading order in $1/N$.
Note that one of the advantages of such an expansion
is that it is a manifestly gauge-invariant expansion ($N$ being
a gauge invariant parameter). It was also observed \cite{HH81}
that it is easier to examine the theory in the limit
$N\to\infty$, where
even the one--loop graviton operator and Faddeev-Popov ghosts
contributions can be neglected with respect to
the contribution coming from a large number N of matter fields.
Thus, in this approximation there are {\it no} corrections from the
quantized degrees of freedom of the gravitational field itself and
gravity can be simply treated classically. Moreover,
observe also that the $N$ fields, although interacting with gravity,
do not couple with another possibly present matter field.
Note also that in the $N\to\infty$ limit, the fluctuations
in the expected stress--energy tensor of the N matter fields
become negligible.
The applicability of the semiclassical field equations,
Eqs.\ (\ref{I.2}) below, is now
given by the higher order $1/N$--corrections,
which suggest that
these equations would (in principle) be valid even
near the Planck scale, where effects which are non
perturbative in $\hbar$ may be important and one
should solve the theory exactly \cite{FW96}.

In this paper we adopt the $1/N$ approach to quantum gravity. 
We then consider the following quadratic
gravitational Lagrangian in four dimensions

\begin{eqnarray}
I=I_{G}+I^{ren}_{m}={1\over 16\pi G}\int d^4x~\sqrt{-g}~
\bigg\{-2\Lambda + R + \alpha R^2 + \beta R_{\mu\nu}
R^{\mu\nu} + 16\pi G{\cal L}^{ren}_m\bigg\}~,
\label{I.1}
\end{eqnarray}
where we have made use of the Gauss--Bonnet invariant to
eliminate the dependence on the Riemann tensor.

The semiclassical field equations derived by extremizing the action $I$
can be written as
\begin{eqnarray}
R_{\mu\nu}-{1\over 2}Rg_{\mu\nu}+\Lambda g_{\mu\nu}+\alpha
\,{}^{(1)}H_{\mu\nu}+\beta\,{}^{(2)}H_{\mu\nu}=
8\pi G\langle T_{\mu\nu}\rangle^{\rm ren}~,\label{I.2}
\end{eqnarray}
where
\begin{eqnarray}
\langle T_{\mu\nu}\rangle^{\rm ren}
\doteq - {16\pi G\over\sqrt{-g}}
{\partial I^{\rm ren}_m\over\partial g^{\mu\nu}}~,  \label{I.2'}
\end{eqnarray}
is the renormalized expectation value of the quantized matter
source and where
\begin{equation}
{}^{(1)}H_{\mu\nu}=-2R_{;\mu\nu}+2g_{\mu\nu}\Box R-
{1\over 2}g_{\mu\nu}R^2+2R R_{\mu\nu}~, \label{I.3}
\end{equation}
and
\begin{eqnarray}
{}^{(2)}H_{\mu\nu}=-2R_{\mu~;\nu\alpha}^{\alpha} +
\Box R_{\mu\nu}+{1\over 2}g_{\mu\nu}\Box R+
2R_{\mu}^{~\alpha}R_{\alpha\nu}-{1\over 2}g_{\mu\nu}
R_{\alpha\beta}R^{\alpha\beta}~. \label{I.4}
\end{eqnarray}
The (renormalized) values of the coupling constants
$\alpha$ and $\beta$ come from the regularization process and
depend on the number and types of fields present as well
as on the method of regularization.
However, since the higher derivative
terms must have very little effect
on the low-frequency domain (in order to avoid conflict with
observations),
is usual to consider them as free numerically small parameters of
the order of $l_{pl}^2$ (or equivalently, $\hbar$, for $G=c=1$).

It was observed
in\ \cite{BSZ85,PS93}, and independently in\ \cite{CLA94},
that one can develop a perturbative approach to that problem
that allows one to write
the perturbative field equations in terms of second order
derivatives of the metric and of the energy--momentum tensor.
The system is thus reduced to an effective Einsteinian one.
This perturbative reduction procedure can be described as follows.
In the perturbative approach\ \cite{CLA94} we consider
coupling constants $\alpha$ and $\beta$ to be of the same order of 
$l_{pl}^2$, and curvatures small 
small enough to ensure that the perturbative series makes
sense
\begin{equation}
|\alpha R|\ll 1~~,~~~~|\beta R_{\mu\nu}|\ll 1~. \label{I.5}
\end{equation}
Note that these conditions arise naturally since the description 
provided by the semiclassical gravity breaks down in a domain where 
the curvature is large i. e. near $|R_{\mu\nu\sigma\delta}|\simeq c^3/
(\hbar G)=3.9\times10^{65}{\rm cm}^{-2}$.

Bearing these conditions in mind, we observe that, up to  
first order in $\alpha$ and $\beta$, the two geometrical conserved tensors 
${}^{(1)}H_{\mu\nu}$ and ${}^{(2)}H_{\mu\nu}$, can be computed in terms of
any known classical zeroth order solution, $g_{\mu\nu}^{(0)}$, 
of Einstein field equations  
\begin{eqnarray}
R_{\mu\nu}^{(0)}-{1\over 2}R^{(0)}g^{(0)}_{\mu\nu}+
\Lambda g^{(0)}_{\mu\nu}={8\pi G\over c^3} T_{\mu\nu}^{(0)}~,
\label{EFE}
\end{eqnarray}
where $T_{\mu\nu}^{(0)}$ is the stress energy momentum tensor of 
some matter fields, which can generally include both 
the source of an ordinary  matter field $F_{\mu\nu}^{(0)}$ 
(e.g. electromagnetic field, perfect fluid, very energetic point--like 
particle, etc), given by 
$T_{\mu\nu}^{O~(0)}=T_{\mu\nu}(g_{\mu\nu}^{(0)}, F_{\mu\nu}^{(0)})$,
and the source of $N$ identical scalar fields not interacting with 
the precedent ordinary matter fields, given by 
$T_{\mu\nu}^{N~(0)}=
T_{\mu\nu}(g_{\mu\nu}^{(0)}, \phi^{(0)~j})$ with $j=1,...,N$. 
All matter fields are then coupled to $g_{\mu\nu}^{(0)}$, so 
that their only relevant back--reaction effects are on the 
gravitational field. Note that since we are interested in 
issues related to the {\it back--reaction} problem, the dynamics 
of the matter fields will not concern us here.  
Our purpose is to solve the equations of motion for gravity
and to find the  back--reaction of the matter fields on the 
gravitational field. 
    
{} From the complete knowledge of the zeroth order classical system, 
i. e.  $(g_{\mu\nu}^{(0)}, F_{\mu\nu}^{(0)}, \phi^{(0)~j})$, 
one can write the `reduced' semiclassical equations to first--order 
(in $\alpha$ and $\beta$) as Einsteinian like ones 
\begin{eqnarray}
R_{\mu\nu}^{(1)}-{1\over 2}R^{(1)}g^{(1)}_{\mu\nu}+
\Lambda g^{(1)}_{\mu\nu}&=&
{8\pi G\over c^3} \langle T_{\mu\nu}^{(1)}\rangle^{\rm ren}
-\alpha{}^{(1)}H_{\mu\nu}(g_{\mu\nu}^{(0)})
-\beta{}^{(2)}H_{\mu\nu}(g_{\mu\nu}^{(0)})\cr\cr
&\doteq& {8\pi G\over c^3} T_{\mu\nu}^{{\rm eff}~(1)}~,
\label{I.6}
\end{eqnarray}
with an  effective source $T_{\mu\nu}^{\rm (1)~eff}$.

Let us now consider the following convenient approximation 
where the renormalized energy--momentum tensor in the expression of 
$T_{\mu\nu}^{\rm (1)~eff}$ can be splitted into two additive parts,   
\begin{eqnarray}
 \langle T_{\mu\nu}^{(1)}\rangle^{\rm ren}
= T_{\mu\nu}^{(1)}+
\langle\Delta T_{\mu\nu}^{N~(1)}\rangle~,\label{I.6'}
\end{eqnarray}
where the first term of the right hand side of this equation includes 
both $T_{\mu\nu}^{O~(1)}$ and $T_{\mu\nu}^{N~(1)}$, and the second term 
includes only the (state--dependent) expected value of the energy--momentum
tensor of the $N$ identical free scalar fields, $\hat{\phi}^{(1)~j}$, 
$\langle\Delta T_{\mu\nu}^{N~(1)}\rangle = 
\langle\Delta T_{\mu\nu}(g_{\mu\nu}^{(1)}, \hat{\phi}^{(1)~j})\rangle =
{\cal O}(l^2_{pl})$. 
A similar approximation can also be found in Ref.\cite{H80}, in the 
linearized version of the one-loop approximation to quantum gravity. 
However, we note that the approximation \ (\ref{I.6'}) is more accurately 
justified in the context of $1/N$ expansion, where at least we have the 
following advantages:
\begin{itemize}

\item The scalar fields $\hat{\phi}^{(1)~j}$ are free (i. e. 
non-self-interacting), so that only one--loop terms in these 
matter fields arise. The stress--energy momentum tensor
of these fields is approximatated in such a way that it 
can conveniently be splitted into a classical part, $T_{\mu\nu}^{N~(1)}$, 
and into a quantum one, $\langle\Delta T_{\mu\nu}^{N~(1)}\rangle$.
 
\item We can use the same argument as that used for the one--loop 
graviton to neglect the one--loop (order $l^2_{pl}$) contribution 
of matter field operator $\hat{F}_{\mu\nu}^{(1)}$, so that the ordinary 
matter, $T_{\mu\nu}^{O~(1)}=T_{\mu\nu}(g_{\mu\nu}^{(1)}, F_{\mu\nu}^{(1)})$,
can always be only treated classically.

\item The state--independent (local) corrections 
are uniquely determined (up the ambiguity given by the value of 
the coupling constants $\alpha$ and $\beta$) by the two terms 
${}^{(1)}H_{\mu\nu}={}^{(1)}H_{\mu\nu}(g_{\mu\nu}^{(0)})$ 
and ${}^{(2)}H_{\mu\nu}={}^{(2)}H_{\mu\nu}(g_{\mu\nu}^{(0)})$, which
can always be build up from the zeroth--order 
classical metric $g_{\mu\nu}^{(0)}$.
 
\end{itemize}

A difficulty still remains. 
The state--dependent part $\langle\Delta T_{\mu\nu}^{N~(1)}\rangle$ requires 
further and independent work to be determined and, 
although we consider here only very simple quantum matter fields, 
to compute this term it is in general a very difficult task.
But things can extremely simplify when the quantum state is build up 
from the conformal vacuum and fields and background metric are 
conformally invariant.
In this situation, in fact, all the corrections can be expressed in
term of local quantities.

Thus, once $T_{\mu\nu}^{\rm (1)~eff}$ is determined, one can try to solve 
the resulting reduced Einstein equations for the metric $g_{\mu\nu}^{(1)}$, 
which in turn will contain the first order corrections in $\alpha$ 
and $\beta$ \cite{CLA94}.  
Note that, formally, the iteration can be pursued up
to the n-th order to find $T_{\mu\nu}^{(n)~eff}$ and 
thus one might solve the n-th order Einstein type field equations to 
find $g_{\mu\nu}^{(n)}$ \cite{CLA95}, 
but since for many purposes it is enough and 
since explicit calculations are very complicated at order higher 
than the first we will stop the process to this order.

A useful procedure, even to the first order in the iteration, is 
to solve for only some of the components of the reduced 
field equations and to enforce the conservation of the effective
energy--momentum tensor, i.e.
\begin{eqnarray}
T^{{\rm eff}~(1)\ ;\nu}_{\mu\nu}=0~.  \label{I.7}
\end{eqnarray}
This is the simplest way  to ensure the {\it  existence} of the 
solutions of the form $g^{(1)}_{\mu\nu}$, without having to solve 
the full set of field equations.

It is evident, from the structure of the process, that we can only 
generate perturbative developments of the solution, if it exists at all.
The knowledge of $g_{\mu\nu}^{(1)}$ allow us to search for 
first quantum corrections to several physical quantities of interest.

It is also evident that for vacuum classical solutions 
\index{Vacuum solution}
the contribution 
of ${}^{(1)}H_{\mu\nu}^{(0)}$ and ${}^{(2)}H_{\mu\nu}^{(0)}$ terms 
vanishes and the only non--zero contribution will be given by the term 
$\langle\Delta T_{\mu\nu}^{N~(1)}\rangle$. 
This happens, for example, in the spacetime of a Schwarzschild 
black hole. Note that from
this, it is also clear now, that the first order solution
to the exterior metric of a gauge string given in
Ref.\ \cite{CLA94}, Eq.\ (23), has the last line wrong.
In fact it gives a $1/r^2$ term outside the core
where $T_{\mu\nu}^{(0)}=0$ and which
is not present in the general relativistic solution.
It is part of the motivations of this paper to correct
this expression.

The rest of the paper is organized as follows:
Sec.\ II\ A. deals with the constant density model for the core
of straight local cosmic strings. We look, first of all, for
static state--independent perturbative solutions to this problem
and we find that they do not exist. We then study in Sec.\ II\ B. 
(local) non--perturbative corrections to the metric by a 
method developed in Ref.\ \cite{AEL93} that makes use of the
properties of the fourth order field equations under conformal
transformations (more generally Legendre transformations).
Sec.\ II\ C. analyses the inclusion of
some particular (non--local) quantum state term in the
energy--momentum tensor that renders possible the existence
of perturbative corrections to the metric.

In Sec.\ III we study the issue
of the existence of equilibrium solutions with spherical
symmetry. Here we also take a constant energy--density $\rho$
while we allow the pressure $p$ to be dependent on the
radial coordinate $r$ to satisfy the Oppenheimer--Volkov
equation. We again show the non consistency of this model
and argue that non--local corrections to the renormalized
energy--momentum tensor hardly can change here the situation.
We end the paper with a further discussion of the
perturbative approach to this uniform--density models.

\section{Cosmic strings with uniform--density cores}

In Refs.\ \cite{BSZ85,PS93} the above perturbative method
have been mainly applied to cosmological models. The
main shortcoming here is the fact that we cannot trust
the results in the most interesting regime; near the initial
singularity. The situation is less dramatic when dealing with
black holes\ \cite{CLA94,CLA95}, since one is mainly interested
in quantities evaluated outside or at the event horizon.
The singularity at $r=0$ being hidden to the outside
world until the very last stages of the black hole
evaporation.

On the other hand, we expect the perturbative approach
to be very precise in the {\it whole} spacetime generated
by a cosmic string (see conditions\ (\ref{I.5})), since
(at GUT scale and below) curvature is moderate inside the
string and vanishes outside.

\subsection{Perturbative solution with local source}

Let us consider a straight (gauge) string lying along the $z$
axis. We have then a cylindrically symmetric system, Lorentz
invariant in the $z$ direction. In this case, a general
enough metric can written as
\begin{equation}
ds^2=e^{2b(r)}(-dt^2+dz^2) + dr^2+r_0^2e^{2a(r)}d\phi^2 ~,
\label{II.2}
\end{equation}
within the following Gaussian coordinates range:
$-\infty<t<+\infty$, $-\infty<z<+\infty$, $0\leq r\leq r_{-}$,
$0\leq\phi<2\pi$.

The non--vanishing components of the Einstein tensor for this
metric read
\begin{equation}
G^r_r=b'(r)\,\left( 2\,a'(r) + b'(r) \right)~,
\label{II.3}
\end{equation}
\begin{equation}
G^\phi_\phi=3\,{{b'(r)}^2} + 2\,b''(r)~,
\label{II.4}
\end{equation}
\begin{equation}
G^t_t={{a'(r)}^2} + a'(r)\,b'(r) + {{b'(r)}^2} + a''(r) +
b''(r)~. \label{II.5}
\end{equation}

Given the energy--momentum tensor (effective or not)
and equating it to the Einstein tensor (over $8\pi$
since we choose units such that $G=c=1$) we
get a set of equations whose solution determine the form
of $a(r)$ and $b(r)$. These equations can be supplemented
with the conservation law which in the case of an
energy--momentum tensor diagonal
and only dependent on the coordinate $r$ (in its
covariant--contravariant form) takes the following
simplified form
\begin{equation}
T_{\mu~;\nu}^\nu=\left[T_{r~,r}^r+2b'(r)
\left(T_r^r-T_t^t\right)+a'(r)
\left(T_r^r-T_\phi^\phi\right)\right]
\delta_\mu^r=0~. \label{II.6}
\end{equation}

The $\phi$--$\phi$ component of Einstein equations can be
reduced to a Riccati equation upon the substitution
$y(r)\dot=b'(r)$. Thus,
\begin{equation}
2y'(r)+3y(r)^2=8\pi T^\phi_\phi~. \label{II.7}
\end{equation}

Once solved this equation for $y(r)$ by making the
sum $G^r_r+G^\phi_\phi-2G^t_t$ we also get a
Riccati equation for $a'(r)\dot=w(r)$
\begin{equation}
w'(r)+w(r)^2=8\pi\,\left[T^t_t-{1\over2}\left(
T^r_r+T^\phi_\phi\right)\right]-y(r)^2~. \label{II.8}
\end{equation}

These two last equations together with the conservation
law\ (\ref{II.6}) are an equivalent set to Einstein
equations\ (\ref{II.3})--(\ref{II.5}) and have the advantage
of being of lower differential order [bearing only first
derivatives in variables $y(r)$ and $w(r)$.]

For the sake of simplicity we will now study the effect of
the state--independent (local) semiclassical corrections to the
uniform density model for the interior structure of straight
cosmic strings.
Thus the only non--vanishing components of
$T^{(0)}_{\mu\nu}$ are\ \cite{G85,L85,H85}
\begin{equation}
T_t^t=T_z^z=-\left({1\over8\pi r_0^2}\right)\Theta(r_s-r)~,
\label{II.9}
\end{equation}
where $r_0$ is a constant that specifies the energy density,
$r_s$ is the ``radius of the string'', and $\Theta$ stands
for the step function.

It is easy to verify that in this case
Eqs.\ (\ref{II.6})--(\ref{II.8}) are satisfied
by\ \cite{G85,L85,H85}
\begin{equation}
b(r)=0
\end{equation}
\begin{equation}
\exp[a(r)]=\cases{\quad
\sin\left({r\over r_0}\right)\quad\quad ;\ r\leq r_s \cr\cr
\left({r\over r_0}\right)\cos\left({r_s\over r_0}\right)
\quad ;\ r\geq  r_0\tan\left({r_s\over r_0}\right)~,}
\label{II.10}
\end{equation}
where we have imposed the equality of the intrinsic metric and
the extrinsic curvature\ \cite{I66} on the matching surface
located at $r_{-}=r_s$ and $r_{+}=r_0\tan(r_s/r_0)$. Thus
$r_s/r_0$
parametrize the family of exterior solutions while $r_0$ gives
the scale of the interior problem.

Given the above expression for the metric and Eq.\ (\ref{II.9})
we can compute the local part of the effective energy--momentum
tensor from Eq.\ (\ref{I.6}). Doing so, we obtain for $r<r_s$
\begin{equation}
T_t^{t~\text {eff}}=T_z^{z~\text {eff}}=-{1\over8\pi r_0^2}+
{(2\alpha+\beta)\over8\pi r_0^4}~,\label{II.11}
\end{equation}
\begin{equation}
T_r^{r~\text {eff}}=T_\phi^{\phi~\text {eff}}=
-{(2\alpha+\beta)\over8\pi r_0^4}~. \label{II.11'}
\end{equation}

Since the local part of effective energy--momentum tensor vanishes
for $r>r_s$, the general form of metric exterior to the string
remain unchanged.
Only the parameter $r_s$ and the constant part of $g_{\phi\phi}$
are expected a priory to vary. We will find, however, a somewhat
unexpected result when we study the internal structure. In fact,
since $T_\phi^{\phi~\text {eff}}=$\ constant, Eq.\ (\ref{II.7}) can
be easily integrated. The general solution reads
\begin{equation}
b(r)={2\over 3}\,\ln\left[\cosh\left(
\sqrt{6\pi \,T_\phi^{\phi~\text {eff}}}\,
\left( r - C_1 \right) \right)\right]+C_2~.
\label{II.12}
\end{equation}

The constants $C_1$ and $C_2$ can be determined from the matching
conditions at $r_s$ with the exterior metric. In particular one
would obtain $C_1=r_s$ and $C_2=0$. Note that since we have
reduced the original fourth order set of equations\ (\ref{I.2})
to an effective one of second order (Eq.\ (\ref{I.6})), the
matching conditions will be that of ordinary General
Relativity\ \cite{I66}:
\begin{eqnarray}
&&b(r)\bigg|^{ext}_{r_{+}}=
b(r)\bigg|^{int}_{r_{-}} ~~{\rm and}~~
a(r)\bigg|^{ext}_{r_{+}}=
a(r)\bigg|^{int}_{r_{-}}
\cr\cr
&&\partial_rb(r)\bigg|^{ext}_{r_{+}}
=\partial_rb(r)\bigg|^{int}_{r_{-}} ~~{\rm and}~~
\partial_ra(r)\bigg|^{ext}_{r_{+}}
=\partial_ra(r)\bigg|^{int}_{r_{-}}~,
\label{MC}
\end{eqnarray}
i. e. the metric and extrinsic curvature should match as the
boundary is approached from each side.

One has also to keep in mind that
since we are considering a perturbative approach, only up to terms
linear in $T_\phi^{\phi~\text {eff}}$ have to be retained
in\ (\ref{II.12}), i.e.
\begin{equation}
b(r)\cong -{(2\alpha+\beta)\over4 r_0^4}(r-C_1)^2+C_2~.
\label{II.13}
\end{equation}

On the other hand, the conservation law\ (\ref{II.6}) applied
to $T_\mu^{\nu~\text {eff}}$
(see Eqs.\ (\ref{II.11})-(\ref{II.11'})) leads to first order to
\begin{equation}
{1\over4\pi r_0^2} b'(r)\cong 0~.
\label{II.14}
\end{equation}

These two last equations are obviously {\it inconsistent} unless
$2\alpha+\beta=0$, what, in our case, leads us back to Einstein
theory. Equivalently, if $b'(r)=0$, Eq.\ (\ref{II.4}) is not
satisfied for the non--vanishing source\ (\ref{II.11}).
What fails? Well, let us recall what our hypothesis have
been. We have supposed that the semiclassical corrections to the
gravitational theory are {\it perturbative} and can be absorbed
in an effective energy--momentum tensor (see Eq.\ (\ref{I.6})).
We then
supposed that the resulting metric could be described by
Eq.\ (\ref{II.2}), cylindrically symmetric, Lorentz invariant and
{\it static}. On this last point let us note that observe in
Eq.\ (\ref{II.12}),
although we have equal and constant energy density and pressure
in the $z$--direction, as in the general relativistic model
that tend to balance each other, it also appears now a constant,
but different from zero, radial and angular component of the
pressure that would destroy the string since they are unbalanced.

But let us keep the condition of staticity and
explore a possible alternative which could
make stable this simple model.

\subsection{Non--perturbative solution with local source}

A second possibility is that the effects of the higher
order terms in the Lagrangian\ (\ref{I.1}) lead to local
non--perturbative corrections. In that case, the method
stated in Sec.\ I would give no result since by its procedure
can only get perturbative corrections. In Ref.\ \cite{AEL93}
it was developed a method to reduce the order of theory from
fourth to second by introducing two auxiliary fields: A scalar
field $\chi$ and a massive spin two field $\psi_{\mu\nu}$.
These fields added to the
usual graviton and account for the extra degrees of freedom
of higher order theories. This method, although in a first
approximation allows to compute non--perturbative corrections
to the general relativistic metric. In fact, in
Ref.\ \cite{AEL93} it was found that (for the sake of
simplicity we take $\beta=0$) the metric solution to the
higher order problem $g_{\mu\nu}^Q$ is conformaly related to
the general relativistic metric $g_{\mu\nu}^E$
\begin{equation}
g_{\mu\nu}^Q\cong\left(1+\chi(r)\right)g_{\mu\nu}^E~,
\label{II.15}
\end{equation}
where the field $\chi$  satisfies the equation
\begin{equation}
({\Box} - m_0^2)\chi =
-{8\pi G\over 3} T^{(Matter)}, \qquad m_0^2\dot={1\over 6\alpha}
\label{II.16}\end{equation}
with the ${\Box}$ operator taken with respect to $g_{\mu\nu}^E$.

The metric outside the core of the string can be
easily found from the above formulae\ \cite{AEL93}
\begin{equation}
\chi(r)=C K_0(m_0 r)~,
\label{II.17}
\end{equation}
where  $m_0$ is real from the no-tachyons constraints
\begin{equation}
3\alpha + \beta\geq 0,~~~~  \beta\leq 0, \label{nt}
\end{equation}
and $C$ is a constant to be determined by the matching
conditions with the internal metric.

To find the internal metric is a little more involved.
We find the solution $\chi(r)=\chi_p(r)+\chi_h(r)$ where
$\chi_p(r)={2\over 3}(r_0^2m_0^2)$ is the particular solution
to\ (\ref{II.16}), and
\begin{equation}
\chi_h(r)''+{1\over r_0}\cot\left({r\over r_0}\right)
\chi_h(r)'-m_0^2\chi_h(r)=0~.
\label{II.18}
\end{equation}

The solution to this equation can be written in terms of
Hypergeometric functions\ \cite{M77}
\begin{eqnarray}
\chi_h(r)&=&C_1F\bigg(a,b, 1, \cos^2\left({r\over2r_0}\right)\bigg)
+C_2\bigg\{F\bigg(a,b, 1, \cos^2\left({r\over2r_0}\right)\bigg)
\ln\left[\cos^2\left({r\over2r_0}\right)\right]
\cr\cr
&&+\sum_{k=1}^{\infty}\cos^{2k}\left({r\over2r_0}\right)
{(a)_k(b)_k\over k^{2}!}
\bigg[\psi(a+k)-\psi(a)+\psi(b+k)-\psi(b)
\cr\cr
&&-2\psi(k+1)+2\psi(1)\bigg]\bigg\}
\label{II.19}
\end{eqnarray}
where coefficients
\begin{eqnarray}
&&a={1+\sqrt{1-{2r_0\over3\alpha}}\over2}
\cr\cr
&&b={1-\sqrt{1-{2r_0\over3\alpha}}\over2}
\end{eqnarray}
shows the essentially non-perturbative character of the solution.
Note that this non perturbative solution is always convergent
(and without poles) in the range of values of the coordinate $r$
and, thus, well behaved at $r=0$, what is in agreement with the
hypothesis made in Ref.\ \cite{AEL93}.

The constants $C$, $C_1$ and $C_2$ have to be determined
with the matching conditions
\begin{eqnarray}
\chi(r)\bigg|^{ext}_{r_{+}}=
\chi(r)\bigg|^{int}_{r_{-}} ~~{\rm and}~~
\partial_r\chi(r)\bigg|^{ext}_{r_{+}}
=\partial_r\chi(r)\bigg|^{int}_{r_{-}}~,
\label{MC2}
\end{eqnarray}
but since both these external and internal non--perturbative
solutions are already of complicate forms
we do not write, here, explicitly their values.

We finally observe, that
although, in Ref.\cite{PS93}, it was emphasized the idea
that a ``self--consistent'' method should be the only valid one
to find ``physical'' solutions of the semiclassical field
equations, in Ref.\cite{CLA94} (where we developed a closely
related approach) we never considered the perturbative method as
an absolute physical prescription for ``throwing out'' all the
non--perturbative solutions. Counterexamples to Ref.\cite{PS93}
claim can, in fact, be found in recent papers\ \cite{CL96,FW96}.

\subsection{Perturbative solution with non--local source}

Other possibility we have to discuss is
to include in the effective energy--momentum tensor a
state--dependent (non--local) piece. The local part of
the total mean value is the one we considered in\ (\ref{I.6}).

It is not difficult to compute $\langle\Delta T_{\mu\nu}\rangle$ outside
the string since the space is locally flat but topologically
and, thus, globally different from Minkowski spacetime. In fact
this non--local part will be the whole correction since the local
part vanishes. One can perform the full quantum field theoretical
computation\ \cite{HK86,L87,FS87} or make simpler symmetry and
conservation arguments\ \cite{H87} to arrive to
\begin{equation}
\langle\Delta T_{\mu\nu}\rangle
={A\hbar\over r^4}\,{\rm diag}(1,1,1,-3)~,
\label{II.20}
\end{equation}
where the constant $A$ depends on the number of conformal
massless free fields we consider and we will keep its value
generic [In Refs.\ \cite{HK86,L87,FS87} it was found that
for a single scalar
field $A=(1440\pi^2)^{-1}[\cos^{-4}(r_s/r_0)-1]$.]

Taking expression\ (\ref{II.20}) for the source of the
semiclassical Einstein equations outside the core of the
string and integrating expressions\ (\ref{II.7}) and (\ref{II.8}),
one gets for the corrected metric up to linear terms in $\hbar$
\begin{equation}
ds^2=\left(1-{4\pi\hbar A\over r^2}\right)\left(-dt^2+dz^2\right)
+dr^2+\left(1+{16\pi\hbar A\over r^2}\right)\cos\left(
{r_s\over r_0}\right)^2r^2d\phi^2~.
\label{II.21}
\end{equation}
with coordinate range:
$-\infty<t<+\infty$, $-\infty<z<+\infty$, $r_+\leq r<+\infty$,
$0\leq\phi<2\pi$, and where now the boundary
reads $r_+=r_0\tan(r_s/r_0)+2\pi A\hbar\cot(r_s/r_0)/r_0$.
The spacetime possess an $r$--dependent deficit angle
\begin{equation}
D=8\pi\mu-16\pi^2 {A\hbar\over r^2}(1-4\mu)
\end{equation}

Now, in the interior of the cosmic string ($r<r_s$), we shall
reverse the problem. Using similar arguments as in
Ref.\ \cite{H87}, we will find the form of the 
$\langle\Delta T_{\mu\nu}\rangle$
inside the string such that a static solution for the (classical)
uniform--density source exists.
Note that although in any static spacetime there exists a wide 
class of Hadamard states\cite{W94} for which this
expectation value makes physical sense, in our
approximation\ (\ref{I.6'}), we chose the $N$ scalar
fields to be in the static vacuum state 
(i. e. $\langle\Delta T_{\mu\nu}\rangle\doteq \langle 0|T_{\mu\nu}^N|0
\rangle$), 
so that these fields do not contribute to the classical matter source
\ (\ref{II.9}). 

Let us first note that when we include the non--local part of
the energy--momentum tensor in the source of Einstein equations,
and consider first order correction in $\alpha$, $\beta$, and
$\hbar$, a formal solution for the metric coefficients in terms
of the source can be easily found. In fact, from\ (\ref{II.7})
and\ (\ref{II.12}) we obtain
\begin{equation}
b(r)\cong C_2-{(2\alpha+\beta)\over4 r_0^4}(r-C_1)^2+4\pi
\int^r dr'\int^{r'}dr''\langle\Delta T_\phi^\phi\rangle~,
\label{II.22}
\end{equation}
while from Eq.\ (\ref{II.8}) we find
\begin{eqnarray}
a(r)&\cong &\ln\left[\sin\left({r\over r_0}\right)\right]+
{C_3\over r_0}-\left[C_4r_0+{(2\alpha+\beta)r\over2r_0^3}
\right]\cot\left({r\over r_0}\right)\nonumber\\
&&-4\pi\int^r{dr'\over
\sin^2\left({r\over r_0}\right)}\int^{r'}dr''\sin^2
\left({r\over r_0}\right)\left(\langle\Delta T^r_r\rangle+
\langle\Delta T^\phi_\phi\rangle-2\langle\Delta T^t_t\rangle\right)~,
\label{II.23}
\end{eqnarray}
where here $\cong$ stands for up to first order terms in
$\alpha$, $\beta$ and $\hbar$. And the constants $C_i$ have
to be determined upon matching this metric with the exterior
one\ (\ref{II.21}).

Again we find that the conservation law\ (\ref{II.6}) proves
to be very useful in analyzing the different possibilities.
Up to first order corrections and in the background of the
constant density string\ (\ref{II.11}) one gets
\begin{equation}
\langle\Delta T^r_r\rangle_{,r}+{\cot\left({r\over r_0}\right)\over r_0}
\left(\langle\Delta T^r_r\rangle-\langle\Delta T^\phi_\phi\rangle\right)+
{b'(r)\over4\pi r_0^2}=0~.
\label{II.24}
\end{equation}

Two cases can be straightforwardly analyzed:
\bigskip
\paragraph{Non--constant radial pressure:}
$\langle\Delta T^r_r\rangle_{,r}\not=0$ and
$\langle\Delta T^r_r\rangle=\langle\Delta T^\phi_\phi\rangle$. 
[Note that this last
equality of components of the energy--momentum tensor is
already fulfilled (c.f.\ (\ref{II.11'}))by the classical and
local one--loop parts of the $T_{\mu\nu}$.]

In this case we have $G_r^r-G_\phi^\phi=0$. From
Eqs.\ (\ref{II.3}) and (\ref{II.4}) we get an equation for
$b(r)$ which can be immediately integrated to give
\begin{equation}
b(r)=-D_1\cos\left({r\over r_0}\right)+D_2~.
\label{II.25}
\end{equation}

The constants $D_1$ and $D_2$ can be determined by the matching
conditions with the exterior metric\ (\ref{II.21}) at $r=r_s$.
Doing so, we obtain
\begin{eqnarray}
D_1&=& {4\pi A\hbar\over r^2_0}{\cot^3\left({r_s\over  
r_0}\right)\over \sin\left({r_s\over r_0}\right)}~,\cr\cr
D_2&=&{2\pi A\hbar\over r^2_0}\cot^2\left({r_s\over r_0}\right)
\left[2\cot^2\left({r_s\over r_0}\right)-1\right]~.
\label{II.26}
\end{eqnarray}

Now, from Eqs.\ (\ref{II.7}) and (\ref{II.11'}) we find
\begin{equation}
\langle\Delta T^r_r\rangle=\langle\Delta T^\phi_\phi\rangle =
{(2\alpha+\beta)\over8\pi r_0^4}+{D_1\over8\pi r_0^2}
\cos\left({r\over r_0}\right)~.
\label{II.27}
\end{equation}

{}From Eq.\ (\ref{II.23}) we see that to find the metric
coefficient $a(r)$ we need also to know
$\langle\Delta T^t_t\rangle$. We can find it making use of the
expression for the trace anomaly of a massless,
conformaly coupled field (see\ \cite{BD82} for
a review on the subject)
\begin{eqnarray}
\langle\Delta T\rangle&=&{\hbar\over2880\pi^2}\left[{\cal A}
C_{\mu\nu\lambda\rho}C^{\mu\nu\lambda\rho}+
{\cal B}\left(R_{\mu\nu}R^{\mu\nu}-{1\over3}R^2\right)+
{\cal C}\Box R\right]\nonumber\\
&=&{\hbar\over2880\pi^2}\left[{\cal A}{4\over3r_0^4}+
{\cal B}{2\over3r_0^4}\right]={\hbar B\over1440\pi^2r_0^4}~.
\label{II.28}
\end{eqnarray}
where ${\cal A}$ and ${\cal B}$ depend on the type
and number of fields considered, and the last equality
defined $B$.

We then find the other two components of the non--local
part of the energy--momentum tensor
\begin{equation}
\langle\Delta T^t_t\rangle=\langle\Delta T^z_z\rangle=
{1\over2}\langle\Delta T\rangle-\langle\Delta T^r_r\rangle~.
\label{II.29}
\end{equation}

We can now explicitly perform the integral in
Eq.\ (\ref{II.23})
\begin{eqnarray}
a(r)&\cong &\ln\left[\sin\left({r\over r_0}\right)\right]+
{D_3\over r_0} +{2\over3}D_1\cos\left({r\over r_0}\right)\cr\cr
&&-\left[D_4r_0-{(2\alpha+\beta)r\over2r_0^3}+
{\hbar Br\over 720\pi r_0^3}\right]\cot\left({r\over r_0}\right)~,
\label{II.30}
\end{eqnarray}
where the constants $D_3$ and $D_4$ can be found from the
matching conditions with the exterior metric\ (\ref{II.21})
at $r=r_s$,
\begin{equation}
D_3=-{2r_0D_1\over3}cos\left({r\over r_0}\right)+
\left[D_4r_0^2-{(2\alpha+\beta)r_s\over2r_0^2}+
{\hbar Br_s\over 720\pi r_0^2}\right]\cot\left({r\over r_0}\right)
+{8\pi A\hbar\over r_0}\cot^2\left({r_s\over r_0}\right)~,
\end{equation}
\begin{eqnarray}
D_4&=&{(2\alpha+\beta)r_s\over2r_0^4}-{\hbar Br_s\over720\pi r_0^2}
+\sin^2\left({r\over r_0}\right)\bigg\{ {2D_1
\sin\left({r\over r_0}\right)\over2r_0}-
{16\pi A\hbar\over r_0^3}\cot^3\left({r_s\over r_0}\right)\cr\cr
&+&\left({\hbar Br_s\over720\pi r_0^2}
-{(2\alpha+\beta)r_s\over2r_0^4}
\right)\cot\left({r\over r_0}\right)\bigg\}~.
\label{II.31}
\end{eqnarray}

\paragraph{Constant radial pressure:}
Let us consider the complementary condition
$\langle\Delta T^r_r\rangle\not=\langle\Delta T^\phi_\phi\rangle$ 
in the simplifying case $\langle\Delta T^r_r\rangle_{,r}=0$.

{}From the conservation law\ (\ref{II.6}) we find that to
first order in $\hbar$
\begin{equation}
y(r)\dot=b'(r)=-4\pi r_0\cot\left({r\over r_0}\right)
\left(\langle\Delta T^r_r\rangle-\langle\Delta T^\phi_\phi\rangle\right)~.
\label{II.32}
\end{equation}

Plugging this equation into\ (\ref{II.7}) we obtain a first
order differential equation for $\langle\Delta T^\phi_\phi\rangle$ which
can be readily integrated to produce
\begin{equation}
\langle\Delta T^\phi_\phi\rangle ={\langle\Delta T^r_r\rangle\over 
\cos^2\left(
{r\over r_0}\right)}+E_1{\sin\left({r\over r_0}\right)
\over \cos^2\left({r\over r_0}\right)}-
{(2\alpha+\beta)\over8\pi r_0^4}\tan\left({r\over r_0}\right)~,
\label{II.33}
\end{equation}
where $E_1$ is a constant of integration.

Now, with this expression we go back to Eq.\ (\ref{II.7}) which
allows us to find
\begin{equation}
b(r)=E_2-{(2\alpha+\beta)r\over 2r_0^3}-4\pi r_0^2\left\{
\langle\Delta T^r_r\rangle\ln\left[\cos\left({r\over r_0}\right)\right]+
E_1\ln\left[
{\cos\left({r\over2r_0}\right)+\sin\left({r\over2r_0}\right)
\over\cos\left({r\over2r_0}\right)-\sin\left({r\over2r_0}\right)}
\right]\right\}~,
\label{II.34}
\end{equation}
where $E_1$ and $E_2$ can be completely determined
by the matching conditions.

{}From the expression for the trace anomaly\ (\ref{II.28})
we find the other two components of the non--local
part of the energy--momentum tensor
\begin{equation}
\langle\Delta T^t_t\rangle=\langle\Delta T^z_z\rangle=
{1\over2}\left[\langle\Delta T\rangle-
\langle\Delta T^r_r\rangle-
\langle\Delta T^\phi_\phi\rangle
\right]~.\label{II.35}
\end{equation}

We can now explicitly perform the integral in
Eq.\ (\ref{II.23})
\begin{eqnarray}
a(r)&\cong&\ln\left[\sin\left({r\over r_0}\right)\right]+
E_2-{(2\alpha+\beta)r\over 2r_0^3}+
{(2\alpha+\beta)\over 2r_0^2}
\ln\left[\cos\left({r\over r_0}\right)\right]
+{E_3\over r_0}\cr\cr
&&-\left[E_4r_0-8\pi{E_0\over r_0}+{(2\alpha+\beta)r\over2r_0^3}+
{\hbar r\over 360\pi r_0^3}\right]\cot\left({r\over r_0}\right)\cr\cr
&&-8\pi E_1\ln\left[{\cos\left({r\over2r_0}\right)+
\sin\left({r\over2r_0}\right)\over\cos\left({r\over2r_0}\right)
-\sin\left({r\over2r_0}\right)}
\right]+{16\pi E_1r_0^2\over\sin\left({r\over r_0}\right)}
\label{II.36}
\end{eqnarray}
The constants $E_3$ and $E_4$ can be then found from the matching
conditions with the exterior metric\ (\ref{II.21}) at $r=r_s$.

To summarize, we have found the explicit form of the expectation value
of the stress--energy--momentum tensor that allow
a consistent static solution for the interior of the straight
cosmic string. 
More explicitly, we have imposed the semiclassical field equations 
and solved them after applying the reduction of order procedure to 
find that the existence of a perturbative solution imply that the 
expectation value of the stress--energy--momentum tensor depends on  
$\alpha$ and $\beta$. Now, we really do not expect the first order 
non--local part to depend on $\alpha$ and $\beta$, since it should 
be computed on the zeroth order background. 
This can be explicitly verified by an independent computation of the 
renormalized energy--momentum tensor using field theory methods, but
this is beyond the scope of the present work.
The form of the expectation value, here, have been chosen {\it ad hoc} 
to satisfy the perturbative static field equations. 
In fact, the dependence on the coupling constants comes
from the imposition of the matching conditions that allow to
determine the integration constants left in the resolution
of the reduced field equations (\ref{II.22})-(\ref{II.24}). 
They contain terms that compensate for the local contributions
to the effective energy--momentum tensor.
Clearly, this is an inconsistency that leaves us with the only option 
of the non--perturbative dependence on $\alpha$ and $\beta$ of the 
straight cosmic string solution.

\section{Spherical stars with constant energy density}

In order to further understand this non existence of
perturbative solutions and see how general this phenomenon might
be, we will study another system with uniform--density, but
this time possessing spherical symmetry instead of a cylindrical
one.

We will represent the spherically symmetric metric by
\begin{equation}
ds^2=-e^{2\Phi(r)}dt^2+e^{2\Lambda(r)}dr^2+r^2
(d\theta^2+\sin^2\theta\,d\varphi^2)~.
\label{III.1}
\end{equation}

The non--vanishing components of the Einstein tensor for the
above metric are
\begin{equation}
G_t^t=-{1\over r^2}\left[r(1-e^{-2\Lambda})\right]'~,
\label{III.2}
\end{equation}
\begin{equation}
G_r^r=-{1\over r^2}(1-e^{-2\Lambda})+{2\over r}\Phi'~,
\label{III.3}
\end{equation}
\begin{equation}
G_\theta^\theta=e^{-2\Lambda}\left[\Phi''+(\Phi')^2+
{\Phi'\over r}-\Phi'\Lambda'-{\Lambda'\over r}\right]~,
\label{III.4}
\end{equation}
\begin{equation}
G_\varphi^\varphi=G_\theta^\theta~.
\label{III.5}
\end{equation}

The Einstein equation derived from equating\ (\ref{III.2}) to
$T_t^t$ can be formally integrated
\begin{equation}
e^{-2\Lambda(r)}=1+{8\pi G\over r}\int_0^rr'^2T_t^t(r')dr'~,
\label{III.6}
\end{equation}
where we have imposed the condition $\Lambda(0)=0$ anticipating
that we are going to consider systems that are will be regular
at the origin of coordinates.

The difference $G_r^r-G_t^t=8\pi G(T_r^r-T_t^t)$ leads to a formal
integral of $\Phi(r)$
\begin{equation}
\Phi(r)=\Phi(0)-\Lambda(r)+{4\pi G}\int_0^rr'e^{2\Lambda(r')}
\left(T_r^r(r')-T_t^t(r')\right)dr'~.
\label{III.7}
\end{equation}

Instead of using the third independent Einstein equation
above, we will consider the conservation of the energy--momentum
tensor equation (which can be derived from the three independent
Einstein equations anyway). The only relevant conservation
equation will be the radial one
\begin{equation}
T_{r,r}^r+{2\over r}\left(T_r^r-T_\theta^\theta\right)+
\left(T_r^r-T_t^t\right)\Phi'=0~.
\label{III.8}
\end{equation}

Combining this last equation with $G_r^r=8\pi GT_r^r$ to
eliminate $\Phi'$ we obtain a constraint on the
energy--momentum components
\begin{equation}
T_{r,r}^r+{2\over r}\left(T_r^r-T_\theta^\theta\right)+
{e^{2\Lambda}\over2r}\left[1-e^{-2\Lambda}
+8\pi GT_r^rr^2\right]\left(T_r^r-T_t^t\right)=0~.
\label{III.9}
\end{equation}

This is the, so called, Oppenheimer--Volkov equation.

Let us next consider a static perfect fluid described by
the following energy--momentum tensor
\begin{equation}
T_t^t=-\rho(r)~,~~T_r^r=T_\theta^\theta=T_\varphi^\varphi=p(r)~,
\label{III.10}
\end{equation}
where $\rho$ is the energy density and $p$ the pressure of an
infinitesimal volume of matter.

An exact solution can be found\cite{S16} when $\rho=\,$constant.
In fact, Eq.\ (\ref{III.6}) can be trivially integrated to give
\begin{equation}
e^{-2\Lambda(r)}=1-{8\pi Gr^2\over 3}~,
\label{III.11}
\end{equation}

It is also easy to integrate the resulting  Oppenheimer--Volkov
equation\ (\ref{III.9})
\begin{equation}
p(r)=\rho\left[{(\rho+3p_c)\sqrt{1-{8\pi G\rho r^2\over 3}}
-\rho-p_c\over 3\rho+3p_c-(\rho+3p_c)\sqrt{1-{8\pi G\rho
r^2\over 3}}}\right]~,
\label{III.12}
\end{equation}
where $p_c=p(0)$. The radial coordinate of the star $R$
is defined as $p(R)=0$, and from the above expression can be
found to be given by
\begin{equation}
R=\sqrt{{3\over8\pi G\rho}\left[1-{\rho+p_c\over\rho+3p_c}
\right]}~,\label{III.13}
\end{equation}
and thus the total mass $M$ of the star will be given by
$M=4\pi G\rho R^3/3$.

To complete the solution, $\Phi$ can be found from
Eq.\ (\ref{III.7})
\begin{equation}
\Phi(r)=\ln\left[{3\over2}\sqrt{1-{8\pi G\rho R^2\over 3}}-
{1\over2}\sqrt{1-{8\pi G\rho r^2\over 3}}\right]~.\label{III.14}
\end{equation}

Now, the above solution will represent our zeroth order
solution in the perturbative scheme of Sec.\ I. In building
up the effective energy--momentum tensor we first observe
that the interior solution we have just described is
conformally flat, i.e. its Weyl tensor vanishes. In this
case tensors ${}^{(1)}H_{\mu\nu}$ and
${}^{(2)}H_{\mu\nu}$ are no longer independent but
related to each other by
${}^{(1)}H_{\mu\nu}=3\,{}^{(2)}H_{\mu\nu}$ and a new
geometrical, object can be defined in a four dimensional
spacetime, which is conserved only in the conformally flat case
\begin{equation}
{}^{(3)}H_{\mu\nu}=-R_\mu^\lambda R_{\nu\lambda}+
{2\over3}RR_{\mu\nu}+{1\over2}R_{\lambda\sigma}
R^{\lambda\sigma}g_{\mu\nu}-{1\over4}R^2g_{\mu\nu}~,
\label{III.15}
\end{equation}

For the background metric\ (\ref{III.11}) and (\ref{III.14})
one obtains\cite{PS93} the first order correction
\begin{equation}
{}^{(3)}H_r^r={}^{(3)}H_\theta^\theta={}^{(3)}H_\varphi^\varphi=
{1\over3}(8\pi G)^2\rho(\rho+2p)\dot={1\over\gamma}\Delta p(r)~,
\label{III.16}
\end{equation}
\begin{equation}
{}^{(3)}H_t^t=-{1\over3}(8\pi G)^2\rho^2\dot=
{1\over\gamma}\Delta\rho~,\label{III.17}
\end{equation}
where $\gamma$ stands for the coupling constant associated
to ${}^{(3)}H_{\mu\nu}$ in the Lagrangian, playing a similar
r\^ole to $\alpha$ and $\beta$.

We first observe that the ${}^{(3)}H_{\mu\nu}$ correction
can be written in a perfect fluid form with the above first
order modifications to the pressure $\Delta p(r)$ and
energy density $\Delta\rho$ (which is still constant throughout
the interior!). However, the bad news for this model are
that these expressions for $\rho+\Delta\rho$ and
$p(r)+\Delta p(r)$ {\it do not} satisfy the
Oppenheimer--Volkov constraint\ (\ref{III.9}), which for a
perfect fluid takes the following simplified form
\begin{equation}
{dp\over dr}=-(\rho+p){e^{2\Lambda}\over2r}\left[
1-e^{-2\Lambda}+8\pi Gpr^2\right]~.
\label{III.18}
\end{equation}

We observe that again we find an incompatibility when we
suppose that perturbative corrections to the
uniform--density model exist. The key point now have been
that the perturbative method determines the form of the
corrected energy--density and the $r$--dependence of the
corrected pressure. We thus no longer have the freedom to
chosse $p(r)$ such to fit the Oppenheimer--Volkov equation.
One can actually check\footnote{These results have been
verified by use of the program of analytic manipulation
``Cartan'', working within ``Mathematica"\cite{W91}.}
that neither the terms coming from
$(\alpha+\beta/3){}^{(1)}H_{\mu\nu}$ corrections are
able to satisfy Eq.\ (\ref{III.9}). In passing we note that
this terms do not generate corrections of the form of a
perfect fluid as did those generated by ${}^{(3)}H_{\mu\nu}$.

Here the non existence of equilibrium solution seems even
more grave than in the case of cosmic strings since we do
not expect the non--local terms in the renormalized
energy--momentum tensor to play any relevant r\^ole. In fact,
the state--dependent part of the vacuum expectation value
of the source vanishes if we suppose that, in addition the
classical fluid, only massless conformally invariant free
fields (such as photons and massless neutrinos) are present.
In this case, in fact, since we are in a conformally flat
spacetime there is no particle production and
the Green's functions can be found by making a conformal
transformation to Minkowski spacetime, finding the appropriate
Green's functions there, and transforming back to the
original spacetime. More complicated coupling can indeed be
considered, but, again, it remains the question of its
physical plausibility.

\section{Conclusion}

Throughout this paper, we have followed the suggestions 
coming from the derivation of an effective action in the 
$1/N$ approximation to quantum gravity in the large $N$ limit. 
We have, thus, assumed that there can exist interesting regimes 
of applicability of a semiclassical theory of gravity.
We estimated that in some physical systems the quantum nature of 
matter is significant compared to the quantum nature of gravity
which might remain negligible.
In fact, although the semiclassical approximation breaks down
at the regime of high (Planck
scale $l_{pl}$) curvature, such as in the final stages of an
evaporating black hole or at very early times in
the evolution of Universe, the effective theory should be
valid in many interesting cases, i. e. when the curvature
approach, but always remains (significantly) less than the
Planck scale.
This is the case of cosmic string and spherical stars where, 
although semiclassical corrections are always small, 
it may happen that these corrections do not allow the interior 
gravitational field to be perturbative in the coupling constants 
$\alpha$ and $\beta$.
In fact, we have seen that if we consider in the back--reaction
problem only local contributions
to the effective source of the semiclassical equations for
uniform--density models of the core of cosmic strings and
relativistic stars, we obtain that the interior metric 
acquires only non--perturbative corrections in the coupling
constants $\alpha$ and $\beta$,
even if the source depends linearly on them.
This shows that one
cannot always simply neglect them or truncate the solutions
to the first perturbative order\ \cite{PS93}.
Note the different
situation here with respect to the case of the ``instability
of the Minkowski space''. While to render Minkowski stable
in Ref.\ \cite{PS93} it was considered the truncation of the
solutions to the perturbative ones and them to first order;
here, we have that such static perturbations do not
exist for the uniform--density models with cylindrical or
spherical symmetry.

We here recall that in the $1/N$ approach to the effective
semiclassical action
it is consistent to consider solutions beyond 
the first perturbative order\ \cite{T77,HH81}.
In fact, in the case of large $N$ limit,
Eqs.\ (\ref{I.2}) are exact,
and therefore it make perfectly sense to look at solutions which
are either perturbatively expandable in powers of $\alpha$
and $\beta$ (to all the orders), as we did for the charged
black hole case in Ref.\ \cite{CLA95} or not--perturbative in those
parameters, as in the present paper.
In this case, it makes also sense to look for exact solutions
to the field equations (See Ref.\ \cite{CL96}) since they 
are obtained from the $1/N$ approximation 
to quantum gravity and they are exact to the leading order 
(in the case of large $N$ limit).
Note also that in the full $1/N$ approximation self--consistency only
requires that these solutions should be of the
same order of the $1/N$ leading order field equations.

One can then consider the contribution
of perturbative, but very particular
non--local terms. This method (plus
considerations of symmetry and the expression of the
trace anomaly) allowed us to determine the state--dependent
terms in the renormalized energy--momentum tensor
that make the static perturbative solution to exist.
What these states correspond to goes beyond the
scope of this paper and should be determined independently
by use of the standard methods of quantum field theory in
curved spacetime. However, one can argue that this
states appear to depend on $\alpha$ and $\beta$ 
already at the first non--trivial order, while we would
expect they only to depend on the zeroth order background
metric.

Finally, we know that the uniform--density model is a first
approximation to the more involved situation inside the
core of strings and neutron stars. In such case of
energy density and pressure dependent on $r$ the analysis
should be possibly made numerically.

\begin{acknowledgments}
The authors benefited from enlighting discussions with
K.Kucha\v{r} and P. H\'aj\'{\i}\v{c}ek.
C.O.L was supported by the NSF grant PHY-95-07719 and by research
founds of the University of Utah.
M.C. holds a scholarship from the Deutscher Akademischer
Austauschdienst.

\end{acknowledgments}


\begin{references}

\bibitem{BD82} N. D. Birrell and P.C.W. Davies, {\it Quantum Fields
in Curved Space}, Cambridge University Press, Cambridge (1982).

\bibitem{BSZ85} L. Bel and H. Serousse--Zia,
Phys. Rev. D, {\bf 32}, 3128 (1985).

\bibitem{PS93} L. Parker and J. Z. Simon, Phys. Rev. D, {\bf 47},
1339 (1993).

\bibitem{CLA94} M. Campanelli, C. O. Lousto and J. Audretsch,
Phys. Rev. D, {\bf 49}, 5188 (1994).

\bibitem{CLA95} M. Campanelli, C. O. Lousto and J. Audretsch,
Phys. Rev. D, {\bf  51}, 6810 (1995).

\bibitem{H80}G. T. Horowitz, Phys. Rev. D, {\bf 21}, 1445 (1980).

\bibitem{AEL93} J. Audretsch, A. Economou and C. O. Lousto,
Phys. Rev. D, {\bf  47}, 3303 (1993).


\bibitem{G85} J. R. Gott III, Astrop. Jour., {\bf 288}, 422 (1985).

\bibitem{L85} B. Linet, Gen. Rel. Grav., {\bf 17}, 1109 (1985).

\bibitem{H85} W. A. Hiscock, Phys. Rev. D, {\bf 31}, 3288 (1985).

\bibitem{I66} W. Israel, Nuovo Cimento B, {\bf 44}, 1 (1966);
ibid {\bf 48}, 463 (1966).

\bibitem{M77}G. M. Murphy, {\it Ordinary Differential Equations
and Their Solutions} (Van Nostrand Co., Princeton, 1960).

\bibitem{CL96} M. Campanelli and C. O. Lousto, preprint gr-qc/9512050,
to appear in {\sl Phys. Rev.} {\bf D 54}, 54 (1996).

\bibitem{FW96} \'E. \'E. Flanagan and R. W. Wald, preprint
gr-qc/9602052.

\bibitem{HK86}T. M. Helliwell and D. A. Konkowski,
Phys. Rev. D, {\bf  34}, 1918 (1986).

\bibitem{L87} B. Linet, Phys. Rev. D, {\bf 35}, 536 (1987).

\bibitem{FS87}V. P. Frolov and E. M. Serebriany,
Phys. Rev. D, {\bf 35}, 3779 (1987).

\bibitem{H87}W. A. Hiscock, Phys. Lett. B, {\bf 188}, 317 (1987).

\bibitem{W94} R. M. Wald, {\sl Quantum field theory in curved spacetime 
and black hole theormodynamics}, Univ. Chicago Press
 (Chicago 1994).

\bibitem{S16}K. Schwarzschild, Sitzber. Deut. Akad. Wiss. Berlin,
424 (1916).

\bibitem{W91}
Calculations have been made aided by S. Wolfram, {\sl Mathematica:
A system for doing Mathematics by Computer}, (Addison-Wesley,
Redwood City,
California, 1991), and the tensor package {\sl Cartan} developed
by H. H. Soleng (1995), gr-qc/9502035.

\bibitem{T77} E. Tomboulis, {\sl Phys. Lett.}
{\bf 70 B}, 361 (1977); {\sl Phys. Lett.}
{\bf 97 B}, 77 (1980).\par

\bibitem{HH81} J. B. Hartle and G. T. Horowitz, {\sl Phys. Rev.}
{\bf D 24}, 257 (1981).\par

\end{references}
\end{document}